\input{psfig.sty}
\documentclass[11pt,a4paper]{article}
\usepackage{jcappub}
\usepackage{epsfig}
\usepackage{color}
\usepackage{subfigure}
\usepackage{psfig}
\usepackage{graphicx}
\usepackage{amssymb}
\usepackage{ulem}

\let\textquotedbl="

\title{A unified solution to the small scale problems of the $\Lambda$CDM model II: introducing parent-satellite interaction}

\author[a,b,c]{A. Del Popolo,}
\author[d]{M. Le Delliou}

\affiliation[a]{Dipartimento di Fisica e Astronomia, University Of Catania, 
Viale Andrea Doria 6, 95125 Catania, Italy}
\affiliation[b]{INFN sezione di Catania, Via S. Sofia 64, I-95123 Catania, Italy}
\affiliation[c]{International Institute of Physics, Universidade Federal do Rio Grande do Norte,\\
59012-970 Natal, Brazil}
\affiliation[d]{Instituto de Fisica Teorica IFT-UNESP, Rua Dr. Bento Teobaldo
Ferraz 271, Bloco 2 - Barra Funda, 01140-070 S\~ao Paulo, SP Brazil}

\emailAdd{adelpopolo@oact.inaf.it, delliou@ift.unesp.br}
\abstract{We continue the study of the impact of baryon physics on the small 
scale problems of the $\Lambda$CDM model, based on a semi-analytical model (Del Popolo, 2009). With
such model, we show how the cusp/core, missing satellite (MSP), Too Big to Fail 
(TBTF) problems and the angular momentum catastrophe can be reconciled with 
observations, adding parent-satellite interaction. Such interaction between dark
matter (DM) and baryons through dynamical friction (DF) can sufficiently flatten
the inner cusp of the density profiles to solve the cusp/core problem. 
Combining, in our model, a Zolotov et al. (2012)-like correction, similarly to 
Brooks et al. (2013), and effects of UV heating and tidal stripping, the number 
of massive, luminous satellites, as seen in the Via Lactea 2 (VL2) subhaloes,
is in agreement with the numbers observed in the MW, thus resolving the MSP and 
TBTF problems. The model also produces a distribution of the angular spin 
parameter and angular momentum in agreement with observations of the dwarfs 
studied by van den Bosch, Burkert, \& Swaters (2001).}
\keywords{cosmology: theory - large scale structure of universe - galaxies:
formation}

\begin{document}
\maketitle

\section{Introduction}

The $\Lambda$CDM (cosmological constant and Cold Dark Matter) model of cosmology, while describing the observations of the Universe, its large scale structure
and evolution very successfully (Spergel et al. 2003, Komatsu et al. 2011; Del Popolo 2007,
2013, 2014a), retains some problems in the description of structures at small
scales (e.g., Moore 1994; Moore et al. 1999; Ostriker \& Steinhardt
2003; Boylan-Kolchin, Bullock, and Kaplinghat 2011, 2012; Oh et al.
2011)%
\footnote{Other remaining problems for the $\Lambda$CDM model involve understanding dark energy: the cosmological constant fine tuning
problem (Weinberg 1989; Astashenok, \& Del Popolo 2012), and the ``cosmic
coincidence problem''.
}. These problems can be enumerated as a) the discrepancy
between cuspy density profiles obtained in N-body simulations
(Navarro, Frenk \& White 1996, 1997 (NFW); Navarro 2010)\footnote{Note that the 
NFW profile, initially considered a universal one, has been shown not to be so 
(e.g., Del Popolo 2010, 2011)} and the flat
profiles of dwarf and Low Surface Brightness galaxies
(Burkert 1995; de Blok, Bosma, \& McGaugh 2003; Del Popolo 2009 (DP09); Cardone
et al. 2011a, 2011b; Cardone \& Del Popolo 2012;
Del Popolo 2012a,b (DP12a, DP12b); Oh et al. 2010, 2011; Kuzio de
Naray \& Kaufmann 2011), coined as the cusp/core problem (hereafter CCP) (Moore
1994; Flores \& Primak 1994;Ogiya \& Mori, 2011,2014; Ogiya et al. 2014), or of Galaxy Clusters (Del Popolo 2014b; Del Popolo \& Gambera 2000); 
b) the discrepancy between the large discs of observed spirals and the small 
discs obtained in Smooth Particle Hydrodynamics (SPH) simulations, referred to 
as the angular momentum catastrophe (AMC, van den Bosch, Burkert,\& Swaters, 
2001); 
c) the discrepancy between the number of predicted and observed subhaloes when 
running N-body simulations (Klypin et al. 1999; Moore et al. 1999)%
\footnote{That difference is larger than an order of magnitude in the Milky Way (MW)!%
}, dubbed the ``missing satellite problem\textquotedbl{} (MSP).

Klypin et al. (1999), and Moore et al. (1999) noticed, in numerical
simulations of galactic and cluster
haloes, an excess of predicted subhaloes compared with observation. They had found $\simeq500$ satellites with circular velocities
larger than Ursa-Minor and Draco, while the MW dwarf Spheroidals (dSphs) are
well known to be far fewer (the Large and Small Magellanic Clouds, and 9 bright dSphs (Boylan-Kolchin, Bullock, and Kaplinghat 2012)).
The problem was later confirmed in subsequent cosmological simulations 
(Aquarius,
Via Lactea II (VL2), and GHALO simulations -- Springel et al. 2008;
Stadel et al. 2009; Diemand et al. 2007). Although insufficiently for a complete solution, it was alleviated with the discovery of the ultra-faint
MW satellites (Willman et al. 2005; Belokurov 2006; Zucker 2006; Sakamoto
\& Hasegawa 2006; Irwin et al. 2007).

The MSP was recently enriched with an extra problem, spawned from the analysis 
of the
Aquarius and the Via Lactea simulations. Simulated haloes produced $\simeq10$
subhaloes (Boylan-Kolchin, Bullock, and Kaplinghat 2011, 2012) that were too
massive and dense to be the host of the MW brightest satellites: while those 
$\Lambda$CDM
simulations predicted in excess of 10 subhaloes with $V_{max}>25$ km/s,
the dSphs of the MW all have $12<V_{max}<25$ km/s. This discrepancy in 
the kinematics
between simulations and the MW brightest dSphs (Boylan-Kolchin,
Bullock, and Kaplinghat 2011, 2012), which is an extra problem of the MSP,
has been dubbed the Too-Big-To-fail (TBTF) problem%
\footnote{``Too big to fail'', in the sense that the extra simulation satellites
 are too
big, compared with MW satellites, to remain invisible.%
} (Ogiya \& Burkert 2014).

Similarly to the solutions to other small scale problems, the resolution of the 
MSP can be classified as either cosmological or astrophysical solutions. 
Cosmological solutions modify either the power spectrum at small scales
(e.g. Zentner \& Bullock 2003), the constituent DM particles
(Colin, Avila-Reese \& Valenzuela 2000; Sommer-Larsen \& Dolgov 2001;
Hu, Barkana \& Gruzinov 2000; Goodman 2000; Peebles 2000; Kaplinghat,
Knox, \& Turner, M. S. 2000) or the gravity theories,
like $f(R)$ (Buchdahl 1970; Starobinsky 1980), $f(T)$ (see Ferraro
2012), and MOND (Milgrom 1983a,b).

Several different kinds of astrophysical solutions have been proposed.
In one picture, the present-day dwarf galaxies could have been more
massive in the past, and they were transformed and reduced to their
present masses 
by strong tidal stripping (e.g., Kravtsov, Gnedin \& Klypin 2004).
Another very popular picture is based on suppression of star formation due
to supernova feedback (SF), photoionization (Okamoto et al. 2008; Brooks et al. 
2013 (B13)), and reionization. In particular, reionization can prevent the
acquisition of gas by DM haloes of small mass, then ``quenching''
star formation after $z\simeq10$ (Bullock, Kravtsov, \& Weinberg
2000; Ricotti \& Gnedin 2005; Moore et al. 2006). This would suppress
dwarfs (dSphs) formation or could make them
invisible. Another solution combines 
the change of central density profiles of satellites from cuspy to
cored (Zolotov et al. 2012 (Z12); B13), which
makes the satellites more subject to tidal stripping and even subject
to being destroyed (Strigari et al. 2007; Pe\~narrubia et al. 2010 (P10)).
Tidal stripping is enhanced if the host halo has a disc. Disc shocking
due to the satellites passing through the
disc produce strong tidal effects on the satellites, even stronger
if the satellite has a cored inner profile. The astrophysical solutions
based on the role of baryons in structure formation, are more easy
to constrain than cosmological solutions, and moreover do not request
one to reject the $\Lambda$CDM paradigm.

While it is not complicated to separately solve the MSP problem, and
the TBTF problem with the recipes discussed above, a simultaneous solution
of both problems in models of galaxy formation based on DM-only simulations
of the $\Lambda$CDM model (Boylan-Kolchin, Bullock \& Kaplinghat 2012)%
\footnote{Note that, in the case of the TBTF problem, the excess of massive 
subhaloes
in MW could disappear if satellites density profiles are modelled through
Einasto's profiles, or if the MW's virial mass is $\simeq8\times10^{11}M_{\odot}$
instead of $\simeq10^{12}M_{\odot}$ (Vera-Ciro et al. 2013; Di Cintio
et al. 2013).
}, is much more complicated.

Previous attempts to find a simultaneous solution to the abundance
problem of satellites (MSP), and to the TBTF problem were made by
the above mentioned Z12, and B13. Z12 found a correction to the velocity
in the central kpc of galaxies, $\Delta v_{{\rm c,1kpc}}$, that mimicked
the flattening of the cusp due to SF and 
tidal stripping.

This correction, together with its subsequent destruction effects from
the tidal field of the baryonic disc, and the identification of subhaloes
that remain dark because of their inefficiency in forming stars due
to UV heating, were then applied by B13 to the subhaloes of the VL2
simulation (Diemand et al. 2008). As a result, the number of massive
subhaloes in the VL2 were brought in line with the number of satellites
of MW and M31.

This work extends a previous paper (Del Popolo {\it et al.} 2014), enriched with
 the part of the model described in appendix \ref{Dynamics of the satellites}, 
and will chiefly focus on the 
latter problem (MSP). However, the model also carries the solution for the 
former two 
(CCP and AMC), from the part of the model developed in Del Popolo {\it et al.} 
(2014). In clear, it uses a semi-analytical model to account for the dynamical 
evolution of satellites. The model, originated in DP09 (and DP12a, b), is an 
improved spherical
infall model already discussed by many authors (Gunn \& Gott
1972; Fillmore \& Goldreich 1984; Bertschinger 1985; Hoffman \& Shaham 1985; 
Ryden \& Gunn 1987; Henriksen \& Widrow 1995, 1997, 1999; Henriksen \& Le 
Delliou 2002; Le Delliou \& Henriksen 2003; Le Delliou 2008; Ascasibar, Yepes \&
 Gottl\"ober 2004; Williams, Babul \& Dalcanton 2004; Le Delliou, Henriksen \& 
MacMillan 2010, 2011a, 2011b)\footnote{Changes to the spherical collapse 
introduced by dark energy where studied in Del Popolo {\it et al.} 2013a; Del 
Popolo, Pace, \& Lima 2013a, b.; Del Popolo et al. 2013b.}.

In the present paper, we follow the path opened by Z12 and B13, but
we consider another mechanism than SF that is also better able to flatten
the density profiles of satellites. Namely, 
we use a mechanism based on the exchange of energy and angular momentum
from baryons clumps to DM through dynamical friction (DF) (El-Zant
et al. 2001, 2004; Ma \& Boylan-Kolchin 2004; Nipoti et al. 2004;
Romano-Diaz et al. 2008, 2009; DP09; Cole et al. 2011; Inoue \& Saitoh
2011). We use DP09 to calculate the flattening of isolated
satellites through the mechanism based on DF. In order to study the effect of 
tidal stripping and heating on the satellites, we use  a combination of the 
procedures from Taylor \& Babul (2001) (TB01) with that 
from P10. Our model differs from the TB01 and P10 models because we use a semi-analytical model based on DF (combination 
coined hereafter TBP model). 
In addition to the difference in the cuspy to cored profile 
mechanism, already present in
B13, our TBP based model is properly taking into account the tidal
heating mechanism. 
Such tidal heating is not captured in the SPH simulations
from which Z12 derive their correction (as stressed in Sect. 4 of B13),
since, as they point out, this would require a very high resolution to be 
captured (Choi et al. 2009).
Moreover, we properly take into account disk shocking while this is neglected in
Z12%
\footnote{B13 is based on Z12 results. In Z12, some haloes experienced disc
shocking and were strongly disrupted. For this, they were considered
outliers, and not used in the calculation in the Z12 correction. %
}: we account for the effects of satellites passing 
through the host galaxy disc. 

Finally, the $\Delta v_{c}$-$v_{{\rm infall}}$ correction that we find shows a
clearer trend (see the discussion in the following section). This is due to the 
absence, in our case, of numerical effects
present, and described, in Z12%
\footnote{The fact that gas-rich satellites in Z12 are too rich is probably
due to inefficient stripping in their SPH simulations. 
} are not present.

In summary, although in our model the profile flattening is calculated as 
in Del Popolo {\it et al.} (2014), here the procedure from the model of Taylor 
\& Babul (2001) is 
included to follow the dynamics of satellites 
and their interactions with the 
main halo, and to take into account the mass loss during substructure evolution 
due to tides and tidal heating (see also Del Popolo \& Gambera 1997). Moreover, inasmuch as inspired by Z12 and B13 
in substructure treatment, we escaped the limitations of their SPH and SPH-based
treatment with semi-analytic methods, obtaining a better $v_{c}-v_{infall}$ 
relation and accessing the effects of tidal heating and disc shocking. Our model
employs a novel combination of parent-satellite interaction through dynamical 
friction, UV heating and tidal stripping to obtain satellite numbers and angular
spin parameter distributions in agreement with observations.

The paper is organised as follows. In Sect. 2, we describe how the
MSP and TBTF problem can be solved simultaneously when baryonic physics
is properly taken into account, extending a model that has been shown to solve 
also the CCP and AMC. 
Appendix \ref{Dynamics of the satellites} gives the detail of the modified 
model, compared with Del Popolo 
{\it et al.} (2014). Sect. 3 describes the results, including a discussion. 
Sect. 4 is devoted to conclusions.

\section{Solving the remaining small scale problems of $\Lambda$CDM}

As we use a model that solves the other problems, we will concentrate here on 
the problems regarding the number and mass of satellites.
Several solutions have been proposed to the
MSP and TBTF problems (Strigari et al. 2007; Simon \& Geha 2007;
Madau et al. 2008; Zolotov et al. 2012; Brooks \& Zolotov 2012; Purcell
\& Zentner 2012; Vera-Ciro et al. 2012; Di Cintio et al. (2012); Wang
et al. 2012). B13 proposed an interesting baryonic solution to those two
problems: instead of running SPH simulations of different galaxies,
they tried to introduce baryonic effects in large N-body dissipationless
simulations, like the VL2, showing that the result obtained is in
agreement with observations of MW and M31 satellites.

In the following, we will partly follow their steps to obtain the
corrected circular velocities and distribution of VL2 satellites.
The differences between our model and Z12 or B13 have been reported
in the introduction.

In summary the method is based on the following ideas and is divided
into two main phases:
\begin{enumerate}
\item In the first phase, the satellite is considered isolated, without
interactions with the host halo, and the flattening of the density profile 
produced by baryonic physics is calculated (in particular,
the lowering of the central mass of subhaloes) in the same fashion as in, e.g., 
Del Popolo {\it et al.} (2014) (see also Hiotelis \& Del Popolo 2006, 2013, to have a semi-analytical description of halos growth.). In this paper we deliberately 
chose not to take account of SF and to concentrate on the model of baryonic 
clumps exchanging energy and angular momentum with DM through DF, since it has 
clearly been shown (Del Popolo, 2014c, Fig. 4) that in the mass (circular 
velocity) range
of the dwarfs studied in the present paper, the former is less efficient in 
transforming cusps into cores than the latter. Del Popolo \& Hiotelis (2014) 
compared also the result of the current model, adding SF, to the SPH simulations
of Inoue \& Saitoh (2012): the full model agrees with the results of Inoue \& 
Saitoh (2012). There, the addition of SF does not alter the outcome 
significantly. This phase is originally described in DP09.

\item Then comes the second phase, when the satellite, no longer considered
isolated, is now subject to the tidal field of the host halo, and
finally accreted to it.
The total central mass is further reduced by tidal stripping and heating. This 
can be expressed in terms of changes in the circular
velocity, $v_{c}$, also proportional to the density. 
More precisely, we calculate the difference in circular velocity,
at 1 kpc, between the DM-only (hereafter DMO) satellites and those
containing also baryons (hereafter DMB satellites), 
$\Delta v_{{\rm c,1kpc}}=v_{{\rm c,DMO}}-v_{{\rm c,DMB}}$, and starting with the same DM
content in our semi-analytical model.

Then, the effects of baryon physics, that are not taken into account
in N-body simulations and responsible of the flattening of the profile,
are introduced in the VL2 simulation 
by correcting the central circular velocity of the satellites, calculating their
mass loss and ascribing them
 stellar masses and luminosities. 


\end{enumerate}
 In other words, we obtain an analytical correction, along a similar idea to 
Z12 but with a semi-analytical model, using tidal stripping and tidal heating 
(recall the latter to be absent in 
Z12) that mimics the effect of flattening of the cusp. To
this we add other corrections (e.g., tidal destruction and UV heating
effects on subhaloes) discussed in DP09 and apply them to the satellites of the 
VL2 simulation, following the same principles as B13.

We stress again that in our model: a) contrary to Z12 and B13, the density 
profile flattening is due to DF and not to SF; b) tidal
heating and disc shocking are taken into account differently from
Z12 and B13; c) our model does not suffer from the numerical effect
producing ``artificially\textquotedbl{} rich satellites in the Z12 simulations.

Since the second phase contains the new method we propose to solve the problems 
involving the number and mass of satellites, we give its description in what 
follows.

\subsection{Mass loss caused by tidal stripping
and tidal heating}

The second phase considers the effects of the interaction between the main halo 
and the satellite.

We follow a combination of TB01 with P10 models' procedures that properly take 
into account 
tidal stripping and heating after infall to extract accurate $v_{max}$ values.
TB01 compared their model with 
high resolution 
simulations, while P10 checked theirs, in their Appendix A, through high 
resolution N-body simulations.

The TB01 model follows the merger history, growth
of the interacting satellites and tracks
the substructure evolution, taking into account the
mass loss due to tidal stripping, tidal heating, as well as 
enhancement of stripping due to the disc, in the host halo. 
The P10 model is fundamentally based on TB01, however not including tidal 
heating.
Our model assumes the same DM host halo NFW (Navarro, Frenk \& White 1996) density profile as P10. It also 
neglects, as P10 and contrary to TB01, the effect of its baryonic bulge. The 
latter assumption is justified by the disc's much larger mass than the bulge's, 
and its 10 times larger density gradient than found in the bulge or the halo.
Such gradient endows the disk with 100 times more heating efficiency on
the satellites than the other components.

This semi-analytic model, indicated as TBP model, is described
in Appendix \ref{Dynamics of the satellites}.

At this point we may put together the mass decrease in satellites
due to phase 1 (core flattening due to interaction of baryonic clumps
with DM), and that due to tidal stripping and heating.

\begin{figure}
\hspace{-1.0cm}
\psfig{file=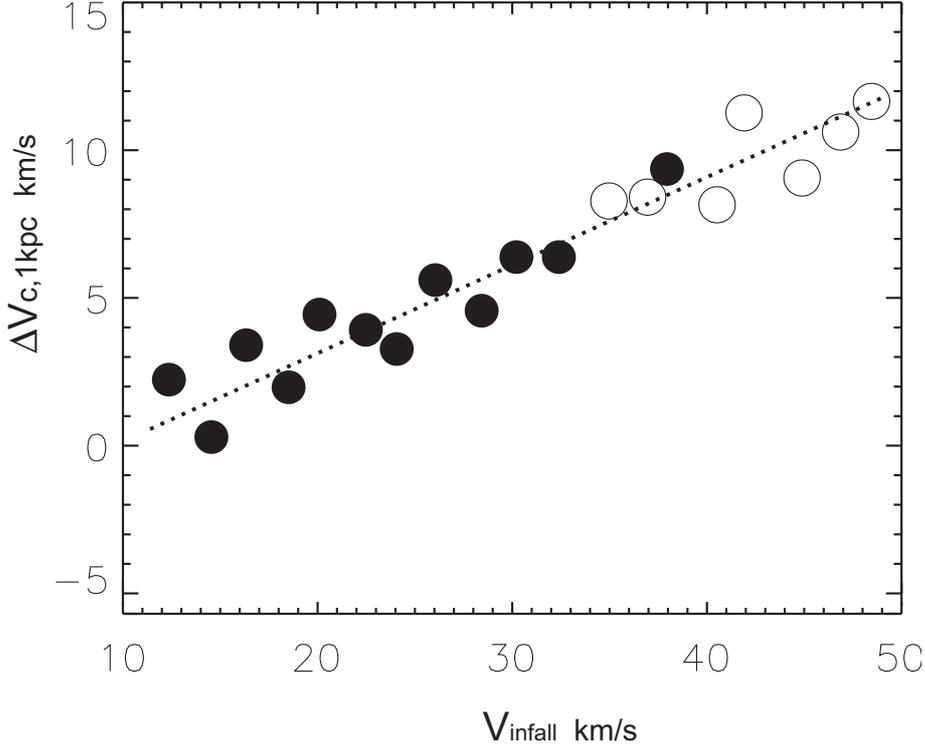,width=13.0cm}
\caption[]{\label{fig:DMO-DMB}Difference in $v_{\rm c}$ at 1 kpc, and at $z=0$, 
between DMO, and DMB 
satellites in terms of $v_{\rm max}$ of the DMO satellites at infall. The filled 
circles correspond to satellites with $M_b/M_{500}<0.01$, while the open circles have 
$M_b/M_{500}>0.01$.}
\end{figure}

In Fig. \ref{fig:DMO-DMB} we plot the difference in circular velocity at 1 kpc, 
and
at $z=0$, between the DMO and DMB satellites of our semi-analytic calculations. 
This difference,
$\Delta v_{{\rm c,1kpc}}=v_{{\rm c,DMO}}-v_{{\rm c,DMB}}$, is due
to the cumulative effects of the two phases of the model:
 a) the flattening from cuspy to cored of density profiles due
to dynamical friction interaction between baryon clumps and DM,
and b) tidal stripping and heating from the crossing of the
satellite through the host galaxy. The dashed line is a fit to the output 
points of the model, and is given by 
\begin{eqnarray}
\Delta(v_{1kpc}) & = & {\rm 0.3v_{infall}-0.3km/s}\nonumber \\
 &  & {\rm 10km/s<v_{infall}<50km/s}\label{eq:mia}
\end{eqnarray}

This correction is then applied to VL2 and is close to the results found by Z12
in the form
\begin{eqnarray}
\Delta(v_{1kpc}) & = & {\rm 0.2v_{infall}-0.26km/s}\nonumber \\
 &  & {\rm 20km/s<v_{infall}<50km/s.}
\end{eqnarray}

They obtained the above equation by fitting the output of their model, based on 
SF and tidal stripping, as displayed on their Fig. 8. Our Fig. \ref{fig:DMO-DMB}
shows a clearer trend $\Delta v_{c}$-$v_{{\rm infall}}$, as it doesn't suffer from 
the numerical effect, described in Z12 (see their Sect. 4 of B13), of 
inefficient stripping in SPH 
simulations.
In addition, Z12 neglects disk shocking, as discussed in the introduction.
These effects, properly taken into account in our model, explain the different 
results we obtain.

Equation \ref{eq:mia}, generated from our semi-analytical model, gives the 
difference between the equivalent of DM and 
enhanced SPH runs, and therefore the corrections to apply to satellites in 
N-body simulations to take account of the missing piece of baryonic physics.

In the case of $v_{infall}={\rm 30km/s}$, the Z12 correction gives
$\Delta(v_{1kpc})=5.74$, while ours gives $\Delta(v_{1kpc})=8.7$.
The difference between the two $\Delta(v_{1kpc})$ 
is due to the different models used to produce the pre-infall flattening
of the satellites density profile and the tidal heating of subhaloes
(Gnedin et al. 1999; Mayer et al. 2001; D'Onghia et al. 2010b; Kazantzidis
et al. 2011).

Indeed, as stated, in Z12 the pre-infall flattening is due to SF, while in our 
case it is 
connected to DF.
As shown by Cole et al. (2011), DF on infalling clumps is a very efficient
mechanism in flattening the DM profile. On one hand, a clump having a mass of 
1\%
of the halo mass can give rise to a core from a cuspy profile, removing
twice its mass from the inner part of the halo. On another hand,  the SF 
mechanism becomes less effective when going to lower masses: dwarfs
 with stellar mass $<10^{5}-10^{7}M_{\odot}$
have fewer stars and thus they contain, as a consequence, less supernovae 
explosions than dwarfs with stellar mass $>10^{7}M_{\odot}$ (Governato 
et al. 2012).

\subsection{Evaluation of luminous satellites}

The previous section exposed how baryonic corrections to dissipationless N-body 
simulations reduce the number of massive satellites. 
We are then left with the task to determine 
whether indeed the baryonic corrections also reduce the number of luminous
satellites that are expected in dissipationless N-body simulations, and in 
particular to the satellites in VL2, and if this 
number is in agreement with those observed
in the MW. In order to check this, other corrections are needed.

Our correction (Eq. \ref{eq:mia}), in the same way as the Z12 correction,
produces satellites that reach $z=0$ with their central
$v_{c}$ reduced by baryonic physics. 
However, some satellites are destroyed (by e.g. stripping or
photo-heating) before $z=0$. 
In N-body simulations, like the VL2, baryonic effects are not taken
into account. In the real universe, or even SPH simulations, enhanced tidal 
stripping (due
to the presence of a disc) may totally destroy some of the satellites seen in 
those N-body simulations. Our method requires then to determine the destroyed 
satellites before applying our Z12-inspired correction to VL2: to evaluate the 
luminous
satellite population, we require the two following corrections: a) to
account for the destruction by tidal stripping, and b) to account
for suppression in star formation. 

\begin{figure}
\hspace{-1.0cm}
\psfig{file=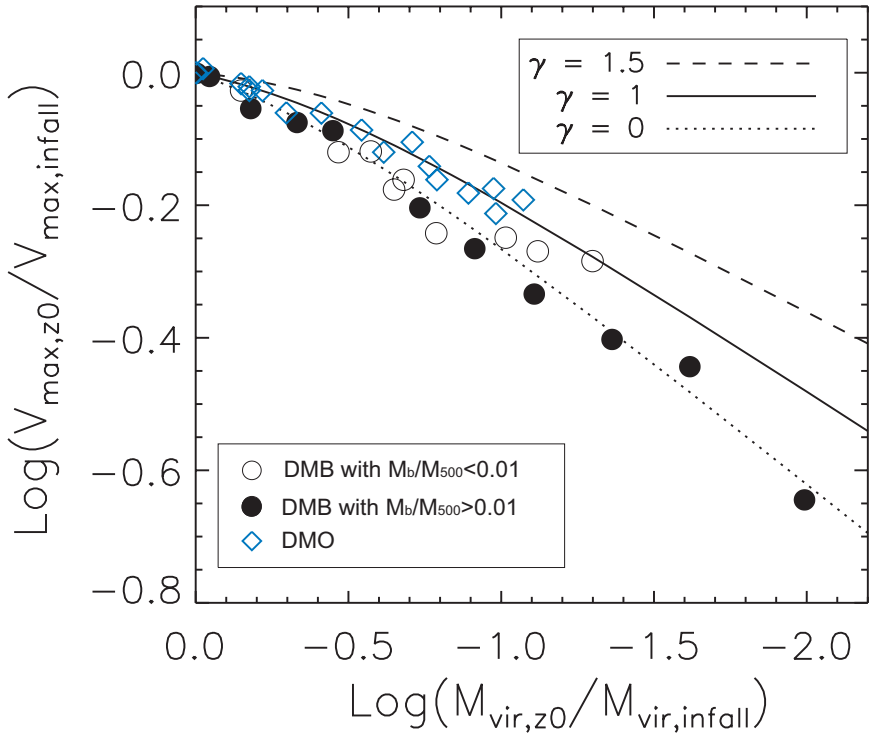,width=13.0cm}
\caption[]{\label{fig:Dv-M}Change in the circular velocity at 1kpc between $z_{\rm infall}$ and 
$z=0$ in terms of the retained mass. Filled 
circles represent the DMB satellites having baryonic fraction $M_b/M_{500}>0.01$,
 while open circles, the DMB satellites with baryonic fraction 
$M_b/M_{500}<0.01$. The open diamonds represent the DMO satellites. The dashed, 
solid, and dotted lines represents Eq. 8 of P10 for slope 
$\gamma=1.5, 1, 0$, respectively. 
}
\end{figure}


The first correction we apply to VL2 N-body satellites is the destruction
rates by tidal stripping. For that,
we need a relation between the mass 
retained since the infall and the change in the velocity (e.g., $v_{{\rm max}}$)
in the same time interval.

We compute that relation applying our semi-analytical model, using the same 
satellites with which we calculated
the relation $\Delta v_{{\rm c,1kpc}}$-$v_{{\rm infall}}$ (see DP09, appendix A, for
the first phase). This population of satellites is split into three groups 
defined below.

\begin{figure}
\hspace{-0.5cm}
\psfig{file=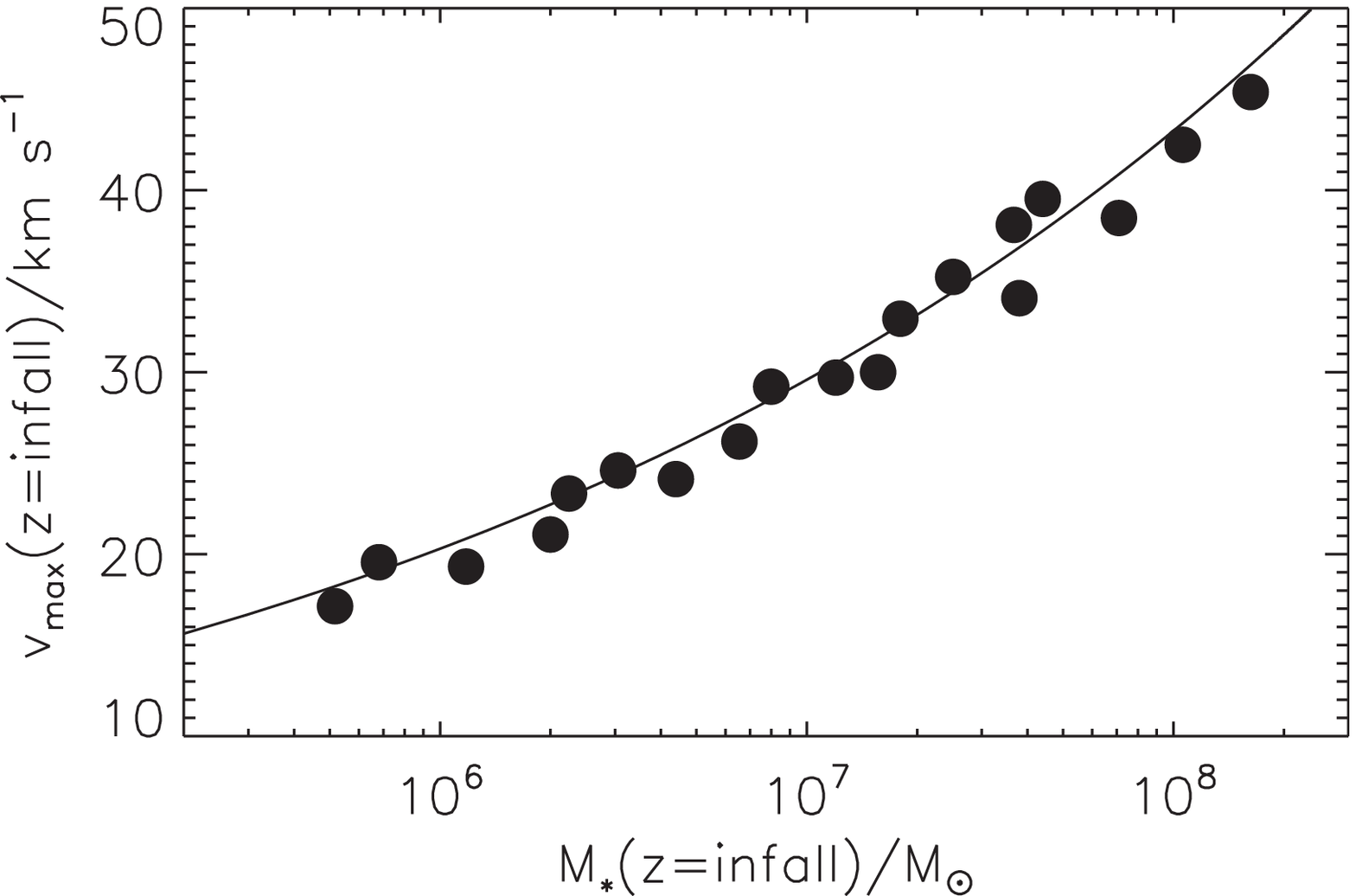,width=13.0cm}
\caption[]{\label{fig:vmax-M}Values of $v_{\rm max}$ of the DMO subhaloes as function of the stellar 
mass, $M_{\ast}$, at infall. The solid line, the fit to the data given by 
$\frac{M_{\ast}}{M_{\odot}}=0.1 (\frac{v_{\rm infall}}{\rm kms^{-1}})^{5.5}$, indicates
how stellar mass change with $v_{\rm infall}$.}
\end{figure}

We plot the result in Fig. \ref{fig:Dv-M}. The filled circles represent the
cored DMB satellites, having baryonic fraction $M_{b}/M_{500}>0.01$, while
the open circles show the cuspy DMB satellites, with baryonic fraction 
$M_{b}/M_{500}<0.01$ (see Governato et al. (2012)).
The open diamonds represent the DMO satellites.

The plot shows that DMB satellites loose more mass than DMOs.
This can be explained by the following reasons: 1) DMB satellites
contain gas, contrary to DMOs; 2) DMB satellites have flatter
profiles than DMOs and thus suffer more tidal stripping (e.g.,
P10). The same goes between the baryon-richer DMB 
(filled circles) and baryon-poorer DMB 
(open circles). 
 The maximum loss happen for DMB satellites in the vicinity of the
host galaxy disc.

In Fig. \ref{fig:Dv-M}, we also plot the analytic fits from Eq. 8 of P10
(see also their Fig. 6), describing the change in $v_{{\rm max}}$
as a function of mass lost due to tidal stripping 
\begin{equation}
\frac{v_{{\rm max}}(z=0)}{v_{{\rm infall}}}=\frac{2^{\zeta}x^{\eta}}{(1+x)^{\zeta}}
\label{eq:pe}
\end{equation}
 where $x\equiv mass(z=0)/mass(z=infall)$.

The dashed line represents the above equation for central density
profile logarithmic slopes $\gamma=1.5$, that corresponds to
$\zeta=0.40$ and $\eta=0.24$, the solid line stands for the case $\gamma=1$, 
for which $\zeta=0.40$ and $\eta=0.30$,
and the dotted line covers the case $\gamma=0$, with $\zeta=0.40$
and $\eta=0.37$, respectively.

The $\gamma=1$ curve in Fig. \ref{fig:Dv-M} gives a good fit to the change in
$v_{{\rm max}}$ for the DMO satellites, and corresponds to cuspy density
profiles. Conversely, the 
$\gamma=0$ curve, that stands for cored profiles, presents a good approximation 
for the DMB satellites, particularly
for those having large baryonic content (i.e., many stars).

Armed with the fit of Eq. (\ref{eq:pe}), we propose to determine the VL2 
satellites that are tidally disrupted
by fixing a destruction criterion (e.g., mass lost). For satellites from N-body 
simulations, such as the VL2,
inner slopes are expected at $\gamma\simeq1$, as found in 
B13. Consequently, 
we fix for them $\zeta=0.40$ and $\eta=0.30$ in Eq. (\ref{eq:pe}).
The fit (\ref{eq:pe}) enables to calculate the mass loss from VL2 satellites' 
velocities at infall and z=0 together with the infall mass. The velocity of VL2 
satellites in the simulation at infall time, $v_{\rm c,VL2,infall}$, is modified 
with Phase 1 correction, $\Delta v_{{\rm c,infall}}$, to account for baryon 
flattening, yielding $v_{{\rm max,infall}}=v_{\rm c,VL2,infall}+\Delta v_{{\rm c,infall}}$. 
Using the correction from Eq. \ref{eq:mia}, the velocity of VL2 satellites in 
the simulation at present $v_{\rm c,VL2,z0}$ is modified into 
$v_{{\rm max,z0}}=v_{\rm c,VL2,z0}+\Delta v_{{\rm c}}$. The infall mass is directly 
obtained from the simulation $M_{\rm vir,infall}=M_{\rm sat,VL2}$.

As for the destruction criteria, we fix it similarly to B13, 
as follows. 
Tides affect much more cored, for which $\gamma=0$,  than cuspy satellites (with $\gamma=1$).
Governato
et al. (2012) found that satellites having a stellar mass $>10^{7}M_{\odot}$,
corresponding to $v_{{\rm infall}}>30$
km/s, are cored, the opposite denoting a cusp. 
Here we assume, as B13 for our cuspy host with a disk and based on Fig. 2 in 
P10, that cored 
satellites, having 
$v_{{\rm infall}}>30$ km/s, are disrupted if they loose 
$>90\%$ of their mass after infall and pass at a distance $<20$ kpc from the
host galaxy centre. In the cases $v_{{\rm infall}}<30$
km/s (cuspy satellite), or cored satellites with pericenters $>20$ kpc, the halo
 is fully stripped off only if it loose 97\%
of its mass (Wetzel \& White 2010).

Summarising,  all the VL2 satellites loosing more than 97\%
mass ($x=0.03$), or loosing
more than 90\% mass, combined with $v_{{\rm infall}}>30$
km/s and a pericentric passages $<20 kpc$, are considered to be destroyed.


The second correction is the suppression of star formation by photo-heating,
obtained from the Okamoto et al. (2008)
results. In their paper, a uniform ionising background is assumed, for which
He II reionization happens at $z=3.5$, while it occurs at $z=9$ for H and 
He I. They found the value of the typical halo mass retaining
50\% of $f_{b}$: $M_{t}(z)$. This mass can be converted
into a typical velocity, $v_{t}(z)$\footnote{In the conversion, we used an overdensity $200\rho_{{\rm crit}}$,
and a WMAP3 cosmology (Spergel et al. 2007).
}.
Thus, if a VL2 subhalo has 
a larger peak velocity,
$v_{{\rm peak}}>v_{t}$\footnote{$v_{{\rm peak}}$ (see Del Popolo \& Gambera 1996 for a definition of the peak mass) represents the largest value of $v_{\rm max}$ over the entire history of the subhalo.}, it is considered
to contain enough baryons to make it luminous. 


The last step consists in assigning a luminosity to the surviving satellites. 
We  first need to allocate stellar masses to VL2 satellites via a relation 
between their $v_{\rm infall}$ and the 
stellar mass $M_\ast$.

To do so, we recycled the pairs of satellites considered in the 
determination of $\Delta v_{\rm c, 1kpc}$.
 DM-only subhaloes are usually associated with their DMB satellites at formation or accretion time (Bullock et al. 2000; Kravtsov et al. 2004;  Strigari et al. 2007;
Bovill \& Ricotti 2011; Simha et al. 2012). Our semi analytical model simply creates  a series of DMO, and a series of corresponding DMB haloes.

Fig. \ref{fig:vmax-M} plots $v_{\rm infall}$ in terms of the stellar mass, $M_\ast$. The 
$v_{\rm infall}$-$M_\ast$ relation is obtained by fitting the data, yielding 
the relation\footnote{Note that tidal stripping and heating from $z_{\rm infall}$ 
to $z=0$ produces a reduction in the halo masses, introducing scatter in the 
$v_{\rm infall}$-$M_\ast$ relation at infall. }

\begin{equation}
\frac{M_{\ast}}{M_{\odot}}=0.1 (\frac{v_{\rm infall}}{\rm kms^{-1}})^{5.5}. 
\end{equation}

Finally, we need to relate $M_\ast$ and the V-band magnitude, $M_{\rm V}$. We 
apply the relation from B13, extracted from Z12 simulations, 
\begin{equation}
\log_{10}(\frac{M_{\ast}}{M_{\odot}})=2.37-0.38 M_{\rm V}.
\end{equation}

\section{Results and discussion}

\begin{figure}
\hspace{-1.0cm}
\psfig{file=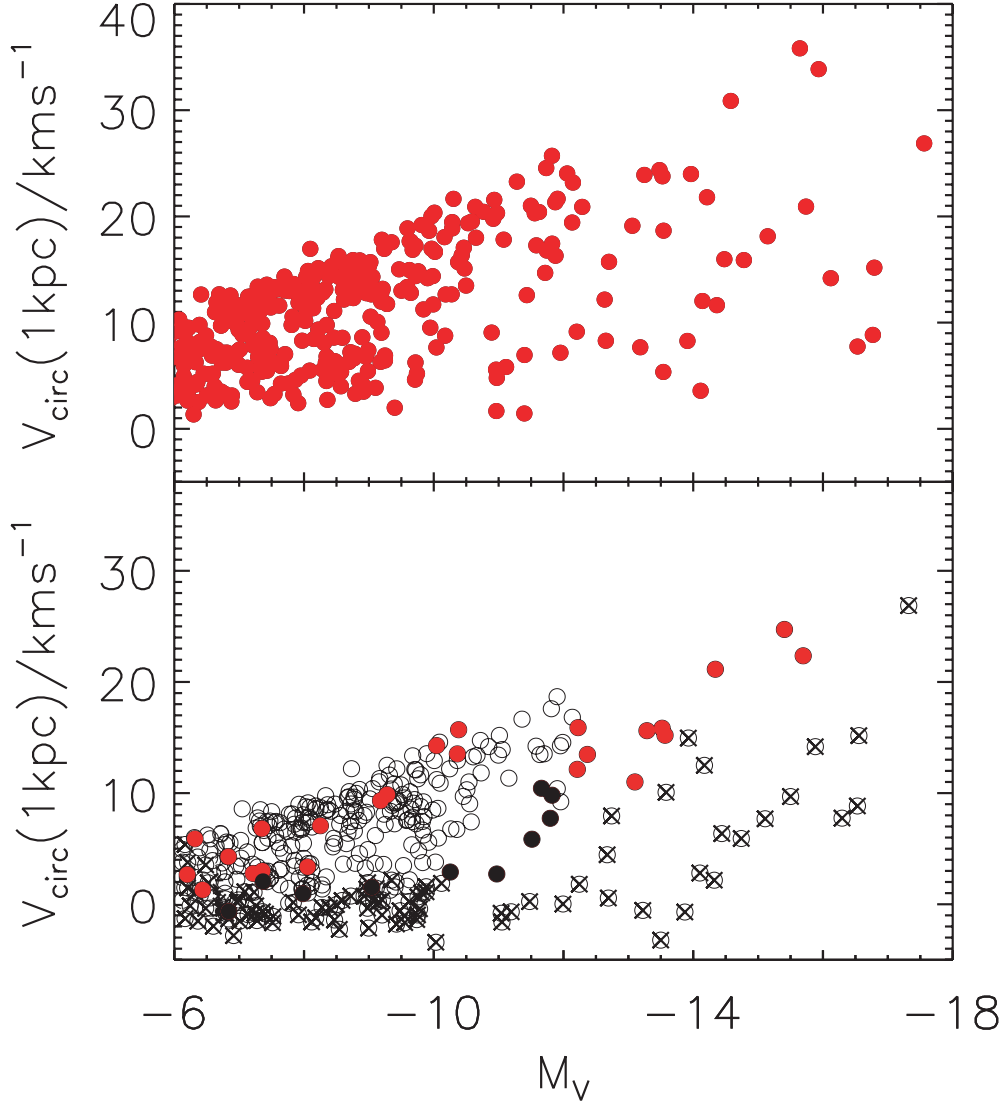,width=13.0cm}
\caption[]{\label{fig:correction}Plot of $v_{1kpc}$ vs $M_{V}$ for the VL2 simulation subhaloes.
In the top panel, we plot the raw VL2 satellites
velocities vs $M_{V}$ at $z=0$, as in
B13. In the bottom panel, we present them after the baryonic corrections 
described in the text. The filled black circles represent satellites that have 
lost enough mass so that their stars are stripped and their luminosities are 
just upper limits, while their actual luminosities are much fainter at infall. 
Filled red circles are satellites actually observable at $z=0$. Dark 
subhaloes are represented by empty circles, while circles with an x are 
subhaloes that have low probability to survive to tidal effects.
}
\end{figure}

The result of the corrections discussed above are plotted in
Fig. \ref{fig:correction}. The top panel represents the raw results from VL2
at $z=0$. The bottom panel presents the results of applying the 
corrections discussed (heating, destruction, and velocity corrections) on the 
same 
satellites. The objects considered ``observable'' in the VL2 simulation are 
ascribed red filled symbols. Filled black circles are satellites that have lost
$\ge$90\% of their mass since infall, but do not satisfy the destruction
criteria previously described: stripped of their stars, they actually 
appear much fainter than the ``observable'' ones (see P10, B13).
Dark objects are marked by empty circles: 
simple empty circles have a mass smaller than the minimum to retain baryon 
and form stars, while objects crossed in addition with an ``x'', represent
subhaloes that do not survive to the baryonic effects (e.g., baryonic
disc, etc).

Note that the Z12 correction was not applied to satellites with 
$v_{max}>{\rm 50km/s}$
(for example, satellites with $M_{V}<-16$, the 5 most
massive satellites at infall of VL2). In fact, those subhalos are 
Magellanic-like and gas-rich at accretion, possibly including an additional 
effect of adiabatic contraction that is not accounted for in the correction. 
The model also assumes small subhalo mass compared to the host. Therefore
Magellanic-types are considered of a different dynamical nature and excluded
from the model, as in, e.g. Simon \& Geha (2007).

We obtain 3 satellites with $v_{1kpc}>20$ km/s, in agreement with B13. 
However, our central velocities are smaller: the correction
to the circular velocity, $\Delta(v_{1kpc})$, is larger in our model
compared to Z12 and B13. In addition, in our case, some satellites
are ``overcorrected'': their corrected velocities are negative.

Similarly to B13, overcorrected haloes are part of a population that lost a 
great part of their mass after
infall. At $z=0$, their circular velocity at 1 kpc, $v_{1kpc}$,
is very low so the correction $\Delta(v_{1kpc})$
brings them to negative values. After infall, that population suffers mass 
loss larger than 99.9\% and exhibit tidal radii $<1$ kpc. It
can therefore be considered as a population of destroyed subhaloes.

It is interesting to note from Fig. \ref{fig:correction} that the model obtains 
not only a 
reduction of the number of satellites, solving the MSP, but also a reduction of
their central velocity, clearing up the TBTF problem.

In analogy with B13, UV heating and tidal destruction are necessary to reconcile
the total number of luminous satellites with observations, while the
Z12-type correction is necessary to reconcile the masses of the subhaloes
with observations.

If the baryonic effects were not taken into account, a population would 
exist of satellites significantly more massive than those of the MW.

Finally, the effect of UV heating is required, in addition to tidal destruction,
 to get the correct number of luminous satellites.


%

In our model, the solution to the aforementioned
problems is connected to the complex interaction between DM and baryons
mediated by DF. Our study is similar to those of El-Zant et al. (2001,
2004), Romano-Diaz et al. (2008), Cole et al. (2011), in the sense
that DF
plays an important role. 
However, while previous studies considered one effect at a time (e.g.,
random angular momentum, angular momentum generated by tidal torques,
adiabatic contraction, cooling, star formation), we 
consider the joint effect of all of them.


Indeed, here the dynamics of the satellites (i.e., the TBP model
in Appendix \ref{Dynamics of the satellites}) proceeds from two competing mechanisms: dynamical friction,
inducing a decay of the satellites orbits, and tidal stripping and
heating, reducing the bound mass of the satellite. This reduction
causes a decrease in the frictional force, which produces in turn
a slowing down of the orbital collapse. Massive and dense satellite
are more subject to DF and sink fast towards the centre of the potential. 
Low-density satellites are more subject to stripping and fall slowly towards 
the centre. Mass loss and tidal heating depend primarily on the satellite 
density profile, as confirmed by P10.


Accounting for tidal heating and disc shocking  speeds up the disruption
of satellite, and yields a further reduction of the mass retained
by them compared with B13. 

\begin{figure}
\hspace{-3.0cm}
\psfig{file=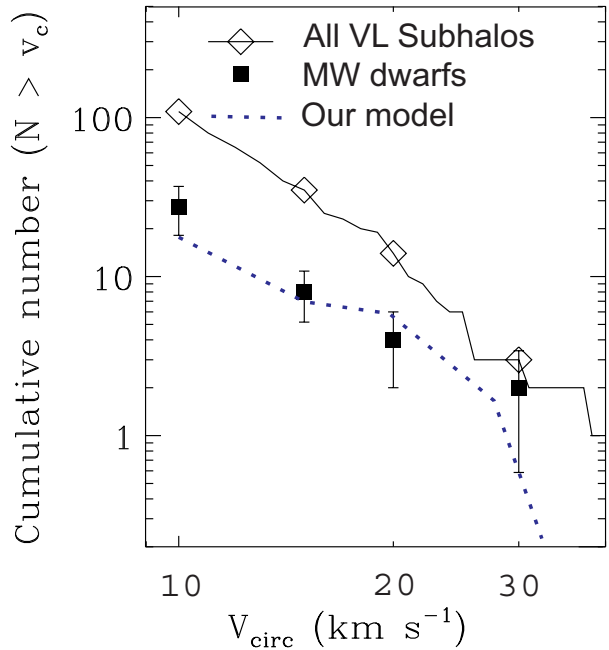,width=16.0cm}
\caption[]{\label{fig:Fn}Cumulative number of MW satellites in terms of circular velocity.
The filled squares display the classical
MW plus ultra-faint-dwarfs in Simon \& Geha (2007). The solid line
with diamonds represents the abundance of
the Via Lactea subhaloes (Diemand et al. 2007). The dashed line shows
the abundance of subhaloes from VL2 after the baryonic corrections
discussed in the text.}
\end{figure}

In Fig. \ref{fig:Fn}, we compare the cumulative number of MW satellites in 
terms
of the circular velocity of the halo with theoretical results. The upper solid 
line with diamonds represents
the Via Lactea subhaloes (Diemand et al. 2007). The filled squares
display the set of the sum of the classical MW dwarfs and the ultra-faint-dwarfs
(Simon \& Geha 2007). The dashed line shows the result of our model
in terms of the abundance of subhaloes in the VL2 simulations after
the baryonic corrections discussed. This figure is built superimposing our 
results, the dotted line, to those of Simon \& Geha (2007) (their Fig. 14) in the
$V_c$ range 10 km/s-40 km/s. This is dictated by a) the fact that 
Eq. \ref{eq:mia} is valid in the range $10< V_{\rm infall}< 50$ km/s, so we 
considered satellites with $V_{\rm infall} >10$ km/s; b) there are no halos at $z=0$
that have circular velocities over 40 km/s. 
This plot demonstrates clearly
how applying the baryonic correction to the VL2 subhaloes reduces
the number of the satellites to reach the levels
observed in the MW, thereby solving
the MSP.

To solve the problem in a single galaxy is not enough to conclude
that the problem is solved in galaxies different from ours. In fact,
several authors have discussed the MSP in relation to the host galaxy
mass. Di Cintio et al. (2012), Vera-Ciro et al. (2012), Wang et al.
(2012), showed that if the MW true virial mass is smaller than $10^{12}M_{\odot}$,
namely $\simeq8\times10^{11}M_{\odot}$, the satellites excess may
disappear. Since our model is not so computationally ``heavy\textquotedbl{}
as SPH simulations, it opens the door to study the MSP in different galaxies.

Summarising, the model shows how taking account of baryon physics allows to 
solve the small scale problems of the $\Lambda$CDM model.

\section{Conclusions}

In the present paper, we looked for a common solution to 
the small scale problems of the $\Lambda$CDM 
using two semi-analytic models: a) the model presented in DP09 (see
also DP12a, b) dealing with isolated satellites, and b) the model based on TB01, and P10 (TBP model) that involves satellite-host dynamics.

The study was divided into two phases: in the first, satellites were 
considered isolated and we studied, by means of
the DP09 model,
how the haloes profile are changed by adiabatic
contraction, dynamical friction and the exchange of angular momentum,
ordered and random, between baryons and DM. This applies both to isolated 
satellites and parent haloes alike, and solves the CCP (Del Popolo {\it et al.} 
2014).

The model had already shown in DP09, DP12a,b, that the angular momentum
generated through tidal torques and random velocities
(random angular momentum) in the system, can be transferred in part to the DM 
from baryons through DF (Del Popolo {\it et al.} 2014). This produces a 
flattening of the cusp
in agreement with previous studies based on DF (El-Zant et al. 2001,
2004; Romano-Diaz et al. 2008; Cole et al. 2011) and SF (Navarro et
al. 1996a; Gelato \& Sommer-Larsen 1999; Read \& Gilmore 2005;
Mashchenko et al. 2006, 2008).

In the second phase, satellites were allowed
to interact with the host halo, and  tidal stripping and
heating were calculated through the TBP model.

We obtained a correction to the central velocity of the satellites
from the cusp to core transformation before the
satellites are accreted, and tidal stripping and heating produced from 
interaction with the main halo. This correction is close to that of Z12.

We then found the relation between the retained mass of satellites
and the changes in $v_{{\rm max}}$ from $z_{{\rm infall}}$ to $z=0$,
and found a connection between mass loss and velocity change, in 
agreement with Eq. 8 of P10. This allowed
us to determine the number of fully disrupted satellites because of
tidal stripping and heating.

This correction, together with the effect of UV heating, and some
criteria to fix which satellites are destroyed by tides, were
applied to the VL2 satellites. As a result, the number of satellites
is reduced and in agreement with the number observed in the MW.
Similarly, the central velocity of satellites is reduced by the
aforementioned corrections, suppressing the angular momentum catastrophe.

The present paper shows that baryonic physics is of fundamental importance
to solve the small scale problems of the $\Lambda$CDM model:
the MSP, the TBTF problem, the CCP (DP09), and the AMC (DP09). The possibility 
to solve those problems
in the $\Lambda$CDM paradigm without the need to change the power
spectrum or the constituent particles of DM is another proof of the
robustness of the $\Lambda$CDM paradigm, and should, in addition,
spur further studies in the direction followed in the
present paper.

\section*{Acknowledgements}

A.D.P. would like to thank the International Institute of Physics in Natal for 
the facilities and hospitality, Adi Zolotov, Alyson Brooks, and Charles Downing 
from Exeter University for a critical reading of the paper. The work of M.Le~D. 
has been supported by FAPESP (2011/24089-5) and PNPD/CAPES20132029. M.Le~D. also
 wishes to acknowledge IFT/UNESP.


\newpage

\appendix

\section{Dynamics of the satellites.}
\label{Dynamics of the satellites}
In the following, we discuss a semi-analytic model that follows the substructure
 evolution within DM haloes. It takes into account the effects of DF, tidal loss 
and tidal heating. The model is basically the TB01 model with small changes 
coming from a similar model by P10. 

Each satellite is represented by a spherically symmetric subhalo, 
whose structure is time dependent. At a time $t$, the satellite's state is 
specified by the form of the density distribution, from a chosen initial 
condition\footnote{The initial density profile of the satellites is given by 
Appendix A.}, by the mass bound to it, and by the heating experienced in time. 
For the determination of the satellite's orbit, we ignore its spatial extent and
 we solve its equation of motion in the potential of the host halo. 

At each time step, the equations solved are:

\begin{equation}
{\ddot {\bf r}}={\bf f}_h +{\bf f}_d + {\bf f}_{\rm df};
\label{eq:eqmot}
\end{equation}
In Eq. (\ref{eq:eqmot}), the term ${\bf f}_h=-GM(<r)/r^2$ is the force due to 
the host halo, where 
\begin{equation}
M(<r)=4\pi\int_0^r \rho(r')r'^2dr';
\label{eq:mass}
\end{equation}
and the density $\rho(r)$ is given by a NFW\footnote{We recall that the NFW 
profile is given by 
$\rho(r)=
\frac{\rho_s}{r/r_s(1+r/r_s)^2}=\frac{\rho_c \delta_v}{r/r_s(1+r/r_s)^2}$, where 
$\delta_v=\frac{\Delta_v}{3} \frac{c^3}{\log(1+c)-c/(1+c)}$ and $\rho_c$ is the 
critical density. The scale radius $r_s$, and $\rho_s$ depend on the formation 
epoch and are correlated with the virial radius of the halo, $R_{\rm vir}$, 
through the concentration parameter $c=R_{\rm vir}/r_s$.} profile with parameters 
$R_{\rm vir}=258$ kpc,
$r_s=21.5$ kpc, $M_{\rm vir}= 10^{12} M_{\odot}$, and $\Delta_v=101$ (Klypin et al. 
2002; Pe\~narrubia et al. 2010).
The term ${\bf f}_{\rm d}$ is the force produced by the baryonic disc. While in 
Pe\~narrubia et al. (2010) it is approximated by means of a Miyamoto-Nagai 
(1975) model, in Klypin et al. (2002) a double-exponential disc is used. We 
select the exponential disc applied in TB01, defined by the density
\begin{equation}
\rho_{\rm d}(r)= \frac{M_{\rm d}}{4 \pi R^2_{\rm d} z_0} exp(-\frac{R}{R_{\rm d}}) 
sech^2(\frac{z}{z_0})
\end{equation}
with $M_{\rm d}= 5.6 \times 10^{10} M_{\odot}$, $r_{\rm d}=3.5$ kpc, and 
$z_0=700$ kpc. 
In this study, we neglect the bulge (similarly to P10), since the disc has a 
much larger mass than the bulge, and presents a steep vertical density gradient.
That gradient is 10 times larger than for the bulge or the halo, resulting
in satellite heating at disc crossing 100 times larger than from the other 
components.

\subsection{Dynamical friction}

The term ${\bf f}_{\rm df}$ is the dynamical friction force on the satellites due 
to the DM particles moving around the host. 
Dynamical friction is approximated through Chandrasekhar's formula 
(Chandrasekhar 1943) which is sufficiently accurate if one can consider the so 
called ``Coulomb logarithm" as a free parameter, fixed through simulations 
(e.g., van den Bosch et al. 1999).  

Chandrasekhar's formula, in our case is given by 
\begin{eqnarray}
 {\bf f}_{\rm df}={\bf f}_{\rm df, disc}+ {\bf f}_{\rm df, halo}= 
 -4\pi G^2 M^2_{sat} \sum_{i=h,d}\rho_i(r) F(<v_{\rm rel,i})\ln \Lambda_i \frac{{\bf 
v}_{\rm rel,i}}{v_{\rm rel,i}^3}.
\label{eq:df}
\end{eqnarray}
having divided the potential into the halo and disc components. $M_{sat}$ is the 
satellite mass, ${\bf r}$ its position, 
and $\ln \Lambda_h$ and $\ln \Lambda_d$ are the Coulomb logarithms of the halo 
and disc components, respectively.    
If ${\bf v}_{sat}$ indicates the velocity vector of the satellite, then 
${\bf v}_{\rm rel,h}={\bf v}_{sat}$ is the satellite's relative velocity with 
respect to the halo, while ${\bf v}_{\rm rel,d}={\bf v}_{sat}-{\bf v}_{d,\phi}$ the 
relative velocity with respect to the disc, while the term 
$v^2_{d,\phi}=R|f_d(Z=0)|$ is the circular velocity of the disc measured on the 
plane of the galaxy. 
The velocity distribution, $F(v)$, is assumed to be isotropic and Maxwellian, 
for simplicity
\begin{equation}
F(<v_{\rm rel,i})={\rm erf}(X_i)-\frac{2X_i}{\sqrt\pi}\exp[-X_i^2];
\label{eq:fv}
\end{equation}
where the term $X_i=|v_{\rm rel,i}|/\sqrt{2}\sigma_i$ is the one-dimensional 
velocity dispersion\footnote{The velocity dispersion is defined as 
$\sigma_i(r)\equiv 1/\rho_i(r)\int_\infty^r \rho_i(r')[f_h(r')+f_d(r')]dr'$.}.

Chandrasekhar's formula was calculated for a massive point particle, but several
 authors showed that it can be applied to calculate the drag force on an 
extended satellite by adjusting appropriately the Coulomb logarithms (e.g., 
Colpi, Mayer \& Governato 1999). Their choice is not trivial. Usually $\Lambda$ 
is defined as $\Lambda=b_{max}/b_{min}$, where 
$b_{max}$ is set to the typical scale of the system, and 
$b_{min} \equiv G(M_{\rm sat}+m)/V^2$, $m$ being the mass of the background 
particles and $V$ the typical velocity of the encounter, is the minimum impact 
parameter.
A different definition is used for an extended satellite (Quinn \& Goodman 
1986). 

The uncertainty in $\Lambda_d$ and $\Lambda_h$ directly reflects on that of the 
orbital decay rates, since the latters depend on the values of the Coulomb 
logarithms. A way to reduce such discrepancy is to treat $\ln \Lambda_h$ and 
$\ln \Lambda_d$ as free parameters. The self-consistent value of the Coulomb 
logarithm best fitting N-body orbits is $\ln\Lambda_h=2.1$ (Pe\~narrubia, Just \& 
Kroupa 2004; Arena \& Bertin 2007), while TB01 and P10 adopt $\ln\Lambda_d=0.5$.
One should also make a correction to the expression for the disc friction, since
 the model assumed a constant satellite wake, and this 
approximation could reveal incorrect if the background density changes over 
small scales (e.g., when the satellite is in the disc plane). This can be corrected
 by smoothing the disc density (see Sec. 2.2.1 of TB01).

\subsection{Mass loss}

A finite size satellite moving through the host galaxy is expected to loose mass
 because of tidal stripping. The mass decrease of the satellite affects its 
dynamic, since the dynamical friction force expression contains $M^2_{\rm sat}$. 
It is clear that we need to estimate the mass loss in order to correctly 
describe the satellite motion. The loss of mass is due to the action of tidal 
forces. 
We
distinguish two model behaviours: if the system is ``slowly varying", we 
consider the material outside a limiting radius, dubbed ``tidal radius", to be 
stripped, while if the system is ``rapidly varying", the satellite material will
 be heated. 

In the first case, one estimates the tidal radius as the distance, measured from
 the centre of the satellite, where the tidal force balances the satellite's 
self-gravity. In the case of satellites on circular orbits, the tidal radius is 
given by
\begin{equation}
R_t\approx \bigg ( \frac{G M_{sat}}{\omega^2 -\rm d^2 \Phi_h/\rm d r^2} 
\bigg)^{1/3};
\label{eq:rt}
\end{equation}
(King 1962), where, as before, $M_{sat}$ is the mass of the satellite $\omega$ is
 its angular velocity, and $\Phi_h$ is the host halo potential. 
Eq. (\ref{eq:rt}) is valid if $ M_{sat}<< M_h$, $R_t<< R_{system}$, and the 
satellite is corotating at $\omega$. 
Eq. (\ref{eq:rt}) describes a steady state loss of mass, while the mass changes 
on a general orbit should depend on the orbital period. One then assumes that 
mass beyond the tidal radius is lost in an orbital period. 

The calculation of ${\rm d}^2\Phi_h /{\rm d}r^2$ is performed averaging over the
 asphericity of the potential originated by the 
disc component, as follows
\begin{equation}
{{{\rm d}^2\Phi_h} \over {{\rm d}r^2}} 
= {{\rm d}\over{{\rm d}r}}\left({{-GM(<r)}\over{r^2}}\right)\,.
\end{equation}

In real systems, satellites are not spherical and do not move inside spherically
 symmetric potentials. In such cases, 
Eq. (\ref{eq:rt}) can be used to define an instantaneous tidal radius. 

The stripping condition can be written in terms of the densities as, 
\begin{equation}\label{overd}
\overline{\rho}_{\rm sat} (< R_{\rm t}) = \xi \overline{\rho}_{\rm gal}(<r)\,.
\end{equation}

The previous equation localises the tidal limit at the radius beyond which the
satellite mean density, $\overline{\rho}_{\rm sat}$, is larger by a factor $\xi$ 
than the average galaxy density inside that radius $r$,  
where
\begin{equation}
\xi \equiv {{\overline{\rho}_{\rm sat} 
(< R_{\rm t})}\over{\overline{\rho}_{\rm gal}(< r)}}= \left({{r^3} \over {G M(< 
r)}}\right) \left(\omega^2 - {{{\rm d}^2\Phi_h} \over {{\rm d}r^2}}\right)
\end{equation}

being $\omega$ the instantaneous angular velocity of the satellite 
and $\omega_c$ is the angular velocity of a circular orbit of radius $r$. 

From the previous discussion, we can define an algorithm to calculate stripping.
 \\
1) We divide the orbital path of the satellite in discrete sections, and 
calculate the tidal radius through 
Eq. (\ref{overd}). \\
2) A fraction $\Delta t / t_{orb}$\footnote{$\Delta t$ is the timestep, while 
$t_{orb} = 2 \pi / \omega$ is the orbital period, which is assumed to be the 
typical time-scale for the mass loss of the satellite.}, of the material outside
 the virial radius will be removed. \\
3)  Whereas in TB01, the satellite was considered disrupted when the tidal 
radius was smaller than the profile core radius, in our case, we define some 
other disruption criteria in Sect. 2.3.

\subsection{Tidal Heating}

As previously discussed, in the case of a rapidly varying gravitational 
potential, shocks are produced 
which result in changes in the satellite structure and give rise to an 
acceleration of the mass loss (e.g., Gnedin \& Ostriker 1997, 1999; Gnedin, 
Hernquist \& Ostriker 1999). A simple first order correction for tidal heating 
can be obtained as follows. Rapid shocks are identified by comparing the orbital
 period of the satellite, $t_{\rm orb,sat}$\footnote{$t_{\rm orb,sat}=2 \pi r_{\rm h} / V_c (r_{\rm h})$ is the 
satellite orbital period at its half-mass radius, $r_h$.}, with the disc shock 
timescale, $t_{\rm shock,d}=Z/V_{Z,{\rm sat}}$.
If $t_{\rm shock,d} < t_{\rm orb,sat}$ the satellite is heated. We then calculate the 
change in energy, and the subsequent mass loss in the satellite. The energy 
change is obtained adopting the impulse approximation (Gnedin, Hernquist, \& 
Ostriker 1999), which yields the velocity change produced by the tidal field in 
the encounter, relative to the centre of the satellite.

This velocity change produced in an encounter of duration $t$, for an element of
 unit mass located at ${\mathbf x}$ with respect to the centre of the satellite,
 writes

\begin{eqnarray}
\Delta {\mathbf V} = \int^t_0 {\mathbf A}_{\rm tid}(t')dt',
\end{eqnarray}
where the term ${\mathbf A}_{\rm tid}$ is the tidal acceleration. 

The first order change in energy is given by
\begin{equation}
\Delta E_1 (t)= W_{\rm tid}(t) = {1 \over 2}\Delta V^2 
\end{equation}

We divide the shock in $n$ time steps of length $\Delta t$ and suppose that the 
satellite is sufficiently small 
so that the tidal acceleration can be expressed in terms of the gradient of the 
gravitational acceleration produced by the external potential, ${\bf g}$. We 
then average the
change of energy on a sphere of radius $r$, in a time step, as
\begin{eqnarray}\label{dW2}
\lefteqn{\Delta W_{\rm tid}(t_n \rightarrow t_{n+1})}\nonumber\\
&=&{1 \over 6}\,r^2\,\Delta t^2\,\Biggr[\,2\,{g}_{a,b}(t_n)\sum^{n-1}_{i = 0} 
{g}_{a,b}(t_i)\nonumber\\
& &\ \ \ \ \ \ \ \ +\ \ \ g_{a,b}(t_n)g_{a,b}(t_n) \Biggr]
\end{eqnarray}
where $g_{a,b} = \partial g_a/\partial x_b$ is evaluated at ${\mathbf x} = 0$%
\footnote{These equations and several others were solved  using the ABS method (Spedicato et al. 2003)
}%
.

The impulse approximation, upon which the calculation of Eq. (\ref{dW2}) is 
based, breaks down in the central part of the 
satellite where the dynamical time-scales can be comparable to, or even shorter 
than, the duration of the shock. When this happens 
the shock effects are significantly reduced.

This is taken into account through a first-order adiabatic correction (Gnedin \& Ostriker 1999)
\begin{equation}\label{adc}
\Delta E_1 = A_1(x)\Delta E_{1,{\rm imp}}\ ,
\end{equation}
where $x=t_{\rm shock}/t_{\rm orb,sat}$ is the adiabatic parameter, and $A_1(x) = (1 +
 x^2)^{-\beta}$, with 
$\beta=5/2$ (Gnedin \& Ostriker 1999).
 
Another correction required is connected to the satellite internal dispersion 
velocity, which is altered by heating (Kundi\'{c} \& Ostriker 1995). We start by
 computing the energy changes at first-order and further take into account the 
higher order effects through the heating coefficient, $\epsilon_{\rm h}$, as   
\begin{equation}
\Delta E = \epsilon_{\rm h} \Delta E_1=\epsilon_{\rm h} A_1(x) \Delta 
E_{1, imp}=\epsilon_{\rm h} A_1(x) \delta W_{tid}.
\end{equation}
Gnedin \& Ostriker (1997) estimated $\epsilon_{\rm h} \simeq 7/3$.

In this paper, we follow TB01 in adopting the value $\epsilon_{\rm h}=3$. 

Practical determination of the effect of heating on the satellite
leads us to assume for each mass element that its potential energy
is proportional to its total energy. We note that shell crossing is
not taken into account by the mass distribution changes.

Consequently, we write that a mass element will have a total energy $E(r)$ 
proportional to $-1/r$, and a radius change 
$\Delta r \propto \Delta E(r)\ r^2$. 

Inside radius $r$ the mean density will change as
\begin{equation}\label{dp}
\Delta \overline{\rho}_r =\ \Delta \left({{3 M(<r)}\over{4 \pi r^3}}\right)\ 
\propto - {\Delta r\over r^4} \propto 
- {\Delta E(r) \over r^2}\ .
\end{equation}

The previous equation shows how the bound mass density in the satellite can 
decrease because of heating, with the results of an acceleration of the mass 
loss. 
The decrease in density can correspond to, either an increase of the velocity 
dispersion in, or an expansion of, the satellite. In any case, it gives rise to 
the same change in the bound mass. 

We can calculate the density change due to tidal heating, at a radius $r$ as a 
function of time. Applying then the equation for tidal stripping, (Eq.\ 
\ref{overd}), to the heated density we can estimate the quantity of mass lost.

In the calculation, we smoothed the disc mass in the vertical direction, as 
already mentioned, over twice the disc scale height. We assume the velocity 
dispersion of the disc to read as
\begin{eqnarray}
\sigma_{\rm h} &=& (V^2_{\rm c,h})^{1/2}/\sqrt{2} ,\hspace{0.7cm}\nonumber\\
{\rm and}\ \ \ \ \ \ \ \ {\phantom{\sigma_{\rm d}}}\nonumber\\
\sigma_{\rm d} &=& V_{\rm c,d}/\sqrt 2 = \sigma_{\rm o}\exp(-R/R_{\rm o})\hspace{0.7cm}
\nonumber
\end{eqnarray}
where $V_{\rm c,h}$, is the circular velocity of the halo, and $V_{\rm c,d}$ that of
 the disc. $\sigma_{\rm o}$ is set to 
$143$ km/s and $R_{\rm o}$ to $7$kpc (namely $2\,R_{\rm d}$), in agreement with 
Velazquez \& White (1999). 

The model depends on three parameters: $\ln \Lambda_{\rm h}$ (strongest 
dependence), $\epsilon_{\rm h}$ (weaker than the previous), and $\ln 
\Lambda_{\rm d}$ (weak dependence). To evaluate the sensitivity of the results to
 parameter variations, 20\% changes in the second parameter ($\epsilon_{\rm h}$) 
were issued: even with such modulations, only  slight changes to the results 
were produced. 
\label{lastpage} 

\end{document}